\documentclass[12pt]{article}
\makeatletter
\def\numberbysection{\@addtoreset{equation}{section}
 	\def\theequation{\thesection.\arabic{equation}}}
\makeatother

\numberbysection

\newcommand{\be}{\begin{eqnarray}}
\newcommand{\ee}{\end{eqnarray}}
\newcommand{\non}{\nonumber}
\newcommand{\kk}{\kappa}

\begin{document}

\begin{titlepage}
\strut\hfill UMTG--235
\vspace{.5in}
\begin{center}

\LARGE Solutions of the boundary Yang-Baxter equation for arbitrary spin\\[1.0in]
\large G. W. Delius\footnote{Department of Mathematics, University of York,
York YO10 5DD, U.K.} 
and Rafael I. Nepomechie\footnote{Physics Department, P.O. Box 248046, 
University of Miami, Coral Gables, FL 33124 USA}\\[0.8in]

\end{center}

\vspace{.5in}

\begin{abstract}
We use boundary quantum group symmetry to obtain recursion formulas
which determine nondiagonal solutions of the boundary Yang-Baxter
equation (reflection equation) of the XXZ type for any spin $j$.
\end{abstract}

\end{titlepage}

\setcounter{footnote}{0}

\section{Introduction}\label{sec:intro}

Solutions $R(u)$ of the Yang-Baxter equation
\be
R_{12}(u-v)\ R_{13}(u)\ R_{23}(v)
= R_{23}(v)\ R_{13}(u)\ R_{12}(u-v) 
\label{YB}
\ee 
play a central role in the study of bulk integrable quantum field
theories and solvable lattice models (see, e.g.,
\cite{ZZ}-\cite{KBI}).  For simplicity we restrict our attention in
this paper to the XXZ case, which is related to the affine Lie algebra
$A^{(1)}_{1}$.  Although early investigations focused on the
fundamental (spin ${1\over 2}$) representation
\be
R^{({1\over 2}\,, {1\over 2})}(u) = \left( \begin{array}{cccc}
	\sinh  (u + \eta) &0            &0           &0            \\
        0                 &\sinh  u     &\sinh \eta  &0            \\
	0                 &\sinh \eta   &\sinh  u    &0            \\
	0                 &0            &0           &\sinh  (u + \eta)
\end{array} \right) 
\label{Rmatrix}
\ee 
(where $\eta$ is the so-called anisotropy parameter), attention soon
turned also to higher-dimensional representations.  The spin $1$ $R$
matrix was obtained \cite{FZ} by direct solution of the Yang-Baxter
equation (\ref{YB}).  A ``fusion'' procedure for $R$ matrices was
subsequently developed in \cite{KRS, KS1}.  However, it was not until
a quantum group approach was formulated by Kulish and Reshetikhin
\cite{KR} that explicit formulas for $R$ matrices of arbitrary spin
became available.  The key feature of this approach is that it
linearizes the problem of finding solutions of the Yang-Baxter
equations.  This work initiated the study of quantum groups (see,
e.g., \cite{Dr, Ji, CP}), and seeded important developments in
integrable quantum field theory (see, e.g., \cite{Za1, BL}).

Solutions $K(u)$ of the boundary Yang-Baxter equation 
\be
R_{12}(u-v)\ K_{1}(u)\ R_{12}(u+v)\ K_{2}(v)
= K_{2}(v)\ R_{12}(u+v)\ K_{1}(u)\ R_{12}(u-v) 
\label{boundaryYB}
\ee 
play a corresponding role for quantum integrable models with boundary
\cite{Ch, Sk, GZ}. The (nondiagonal) fundamental representation
\be
K^{({1\over 2})}(u) = \left( \begin{array}{cc}
\sinh(\xi + u)   & \kk \sinh  2u \\
\kk \sinh  2u     & \sinh(\xi - u) 
\end{array} \right) 
\label{Kmatrix}
\ee 
(where $\xi$ and $\kk$ are boundary parameters) was found \cite{GZ,
dVGR} by direct solution of (\ref{boundaryYB}), as was the spin $1$
result \cite{IOZ}.  A fusion procedure for $K$ matrices was developed
in \cite{MNR, MN1, Zh}.  However, the problem of finding explicit
formulas for $K$ matrices of arbitrary spin has so far remained
unsolved.  Some partial results include work on the so-called
reflection algebra \cite{KS2, KuSa} and on Liouville theory for open
strings \cite{CG}.

We present here some further progress on this problem.  Namely, we
obtain recursion formulas which determine the matrix elements of
$K^{(j)}(u)$ for any spin $j$.  Our approach, generalizing the one
used to solve the corresponding bulk problem \cite{KR}, is based on
``boundary quantum groups'' \cite{MN2, DM}.  One application of this
result is to determine certain coefficients appearing in the Bethe
Ansatz solution \cite{Ne} of the open XXZ quantum spin chain with
nondiagonal boundary terms at roots of unity.

In Section \ref{sec:bulk} we review the construction of the quantum
group generators which commute with the $R$ matrix (\ref{Rmatrix}) and
its higher-spin generalization.  In Section \ref{sec:boundary} we first
recall \cite{MN2} the combinations of these quantum group generators
which commute with the $K$ matrix (\ref{Kmatrix}). By demanding that 
these same combinations of generators also commute with $K^{(j)}(u)$,
we obtain a set of linear relations, which we then solve for the 
matrix elements $K^{(j)}_{mn}(u)$ .  In Section \ref{sec:app} we
apply these results to the problem of the open XXZ quantum spin chain
with nondiagonal boundary terms at roots of unity. We end with
a brief discussion in Section \ref{sec:disc}.

\section{The bulk case}\label{sec:bulk}

In this Section, we review the construction \cite{KR, BL} of the
quantum group generators which commute with the $R$ matrix
(\ref{Rmatrix}) and its higher-spin generalization. To this end,
it is convenient to introduce the two-component Faddeev-Zamolodchikov
``particle-creation operators'' 
\be
A(u) = \left( \begin{array}{l}
             A_{+}(u) \\
             A_{-}(u) 
             \end{array} \right) \,, \non 
\ee
in terms of which the $R$ matrix $R^{({1\over 2}\,, {1\over 2})}(u)$ 
can be defined by
\be
A(u_{1}) \otimes A(u_{2}) =
\check R^{({1\over 2}\,, {1\over 2})}(u)\ 
A(u_{2}) \otimes A(u_{1}) \,,
\ee
where $\check R^{({1\over 2}\,, {1\over 2})} 
={\cal P} R^{({1\over 2}\,, {1\over 2})}$, 
${\cal P}$ is the permutation matrix, and $u = u_{1} - u_{2}$. 
Associativity of the products in 
$A(u_{1}) \otimes A(u_{2}) \otimes A(u_{3})$
then leads \cite{ZZ} to the Yang-Baxter equation (\ref{YB}).

Let us assume the following commutation relations of the quantum group
generators $Q_{\pm}$, $\bar Q_{\pm}$ and $T$ with the
particle-creation operators
\be
Q_\pm A(u) &=&  q^{\pm \sigma_3} A(u) 
Q_\pm +   e^{u} \sigma_\mp A(u) 
\,,  \non \\         
\bar Q_\pm A(u) &=&  q^{\mp \sigma_3} A(u) 
\bar Q_\pm +  e^{-u} \sigma_\mp  A(u)  \,,  \non \\
T A(u)  &=& A(u)  T + \sigma_3 A(u)  
\,, \label{comrel}
\ee
where 
\be
q=e^{\eta} \,,
\ee
and $\sigma_{\pm}= {1\over 2}(\sigma_{1} \pm i \sigma_{2})$. 
Associativity of the products in $Q A(u_1)  \otimes 
A(u_2)  |0\rangle$ and invariance of the vacuum 
$Q |0\rangle =0$ (where $Q = Q_\pm \,, T$ or $Q = \bar Q_\pm  \,, T$ ); 
or, equivalently,
\be
\left[ \check R^{({1\over 2}\,, {1\over 2})}(u) \,, \Delta (Q) \right] = 0
\ee
(where $\Delta$ is the comultiplication) leads to the $R$ matrix
(\ref{Rmatrix}).

This construction generalizes to arbitrary spin 
$j \in \{ {1\over 2}\,, 1 \,, {3\over 2}\,, \ldots \}$. We introduce
the $(2j+1)$-component particle-creation operators 
$\tilde A(u) $, in terms of which the $R$ matrix 
$R^{({1\over 2}\,, j)}(u)$ can be defined by
\be
A(u_{1})  \otimes \tilde A(u_{2})  =
\check R^{({1\over 2}\,, j)}(u)\ 
\tilde A(u_{2})  \otimes A(u_{1})  \,.
\ee
where $\check R^{({1\over 2}\,, j)} = R^{({1\over 2}\,, j)} P$, and
$P$ is a $2(2j+1) \times 2(2j+1)$ matrix which satisfies 
\be
P \left( \tilde M \otimes N \right) P^{-1} = N \otimes \tilde M \,,
\ee
where $\tilde M$ and $N$ are arbitrary $(2j+1) \times (2j+1)$ and 
$2 \times 2$ matrices, respectively.

We assume the commutation relations
\be
Q_\pm \tilde A(u)  &=&  q^{\pm 2 H} \tilde A(u)  
Q_\pm +   e^{u+{\eta\over 2}} q^{\pm H} E_{\mp} \tilde A(u)  
\,,  \non \\         
\bar Q_\pm \tilde A(u)  &=&  q^{\mp 2 H} \tilde A(u)  
\bar Q_\pm +  e^{-u-{\eta\over 2}} q^{\mp H} E_{\mp}  \tilde A(u)  
\,,  \non \\
T \tilde A(u)  &=& \tilde A(u)  T + 2 H \tilde A(u)  
\,, \label{comrelj}
\ee
where the matrices $H$ and $E_{\pm}$ have matrix elements
\be
(H)_{mn} &=& (j+1-n) \delta_{m,n} \,, \qquad m \,, n = 1 \,, 2 \,, 
\ldots \,, 2j+1 \,, \non \\
(E_{+})_{mn} &=&  \omega_{m}\delta_{m,n-1} \,, \qquad
(E_{-})_{mn} = \omega_{n}\delta_{m-1,n} \,, \qquad 
\omega_{n}=\sqrt{[n]_{q}\ [2j+1-n]_{q}} \,,
\label{matrixelements}
\ee
and 
\be
[x]_{q} = {q^{x} - q^{-x}\over q - q^{-1}}  \,.
\ee
These matrices form a $(2j+1)$-dimensional representation of the
$U_{q}(su(2))$ algebra
\be
[  H \,,  E_{\pm} ] = \pm  E_{\pm} \,, \qquad 
[ E_{+} \,, E_{-} ] = [2 H]_{q} \,.
\ee
For $j={1\over 2}$, the relations (\ref{comrelj}) reduce to (\ref{comrel}).
Associativity of the products in $Q A(u_1)  \otimes 
\tilde A(u_2)  |0\rangle$ and invariance of the vacuum 
$Q |0\rangle =0$ (where $Q = Q_\pm \,, T$ or $Q = \bar Q_\pm  \,, T$ )
leads to the $R$ matrix \cite{KR}
\be
R^{({1\over 2}\,, j)}(u) = \sinh (\eta) \left(\sigma_{+} \otimes E_{-} + 
\sigma_{-} \otimes E_{+} \right) 
+ \sinh \left(u + \left({1\over 2}+\sigma_{3} \otimes H\right) \eta \right) 
\,.
\label{Rmatrixj}
\ee

\section{The boundary case}\label{sec:boundary}

Having reviewed the construction of the quantum group generators which
commute with the $R$ matrix, we now turn to the boundary case. The 
matrix $K^{({1\over 2})}(u)$ can be defined by \cite{GZ}
\be
A(u)  |0\rangle_{B} = K^{({1\over 2})}(u) A(-u)  |0\rangle_{B} \,,
\ee
where $|0\rangle_{B}$ is the vacuum (ground state) in the boundary
case.  Associativity of the products in $A(u_{1}) \otimes A(u_{2})
|0\rangle_{B}$ then leads \cite{GZ} to the boundary Yang-Baxter
equation (\ref{boundaryYB}).

Following \cite{MN2}, we consider the combinations of 
quantum group generators 
\be
\hat Q &=& \bar Q_{+} + Q_{-} - {e^{-\xi}\over 2 \kk \sinh \eta}q^{-T}
\,, \non \\
\hat Q' &=& \bar Q_{-} + Q_{+} + {e^{\xi}\over 2 \kk \sinh \eta}q^{T}
\,,
\label{boundaryqg}
\ee
which generate the boundary quantum group. Indeed, associativity of 
the products in $\hat Q A(u)  |0\rangle_{B}$ and 
$\hat Q' A(u)  |0\rangle_{B}$, together with invariance of the vacuum 
$Q_{\pm} |0\rangle_{B} = \bar Q_{\pm} |0\rangle_{B} = T |0\rangle_{B} =0$, 
imply (using the commutation relations (\ref{comrel})) the $K$ 
matrix (\ref{Kmatrix}).

The spin $j$ matrix $K^{(j)}(u)$ can similarly be defined by 
\be
\tilde A(u)  |0\rangle_{B} = K^{(j)}(u) \tilde A(-u)  |0\rangle_{B} \,.
\ee
Associativity of the products in $\hat Q \tilde A(u)  |0\rangle_{B}$ and
$\hat Q' \tilde A(u) |0\rangle_{B}$ and invariance of the vacuum
imply (using the commutation relations (\ref{comrelj})) 
\be
\lefteqn{\left(e^{-u-{\eta\over 2}}q^{-H}E_{-} + e^{u+{\eta\over 2}} q^{-H} E_{+}
- {e^{-\xi}\over 2 \kk \sinh \eta}q^{-2 H} \right)  K^{(j)}(u)} \non \\ 
&=& K^{(j)}(u) \left(e^{u-{\eta\over 2}}q^{-H} E_{-} 
+ e^{-u+{\eta\over 2}}q^{-H} E_{+}
- {e^{-\xi}\over 2 \kk \sinh \eta}q^{-2 H} \right) 
\ee
and
\be
\lefteqn{\left(e^{-u-{\eta\over 2}}q^{H}E_{+} + e^{u+{\eta\over 2}} q^{H} E_{-}
+ {e^{\xi}\over 2 \kk \sinh \eta}q^{2 H} \right)  K^{(j)}(u)} \non \\ 
&=& K^{(j)}(u) \left(e^{u-{\eta\over 2}}q^{H} E_{+} 
+ e^{-u+{\eta\over 2}} q^{H} E_{-}
+ {e^{\xi}\over 2 \kk \sinh \eta}q^{2 H} \right) \,,
\ee
respectively.  Making use of the explicit expressions
(\ref{matrixelements}) for the matrix elements of $H$ and $E_{\pm}$,
we obtain the relations
\be
\lefteqn{e^{-u-\eta(j+{3\over 2} -m)} \omega_{m-1} K^{(j)}_{m-1 \,, n}(u) +
e^{u-\eta(j+{1\over 2} -m)} \omega_{m} K^{(j)}_{m+1 \,, n}(u)
- {e^{-\xi}\over 2 \kk \sinh \eta} e^{-2 \eta (j+1-m)}K^{(j)}_{m n}(u)} 
\non \\
&=& 
e^{u-\eta(j+{1\over 2} -n)} \omega_{n} K^{(j)}_{m \,, n+1}(u) +
e^{-u-\eta(j+{3\over 2} -n)} \omega_{n-1} K^{(j)}_{m \,, n-1}(u)
- {e^{-\xi}\over 2 \kk \sinh \eta} e^{-2 \eta (j+1-n)}K^{(j)}_{m n}(u) \,,
\non \\
\label{rel1}
\ee
and
\be
\lefteqn{e^{-u+\eta(j+{1\over 2} -m)} \omega_{m} K^{(j)}_{m+1 \,, n}(u) +
e^{u+\eta(j+{3\over 2} -m)} \omega_{m-1} K^{(j)}_{m-1 \,, n}(u)
+ {e^{\xi}\over 2 \kk \sinh \eta} e^{2 \eta (j+1-m)}K^{(j)}_{m n}(u)} 
\non \\
&=&
e^{u+\eta(j+{3\over 2} -n)} \omega_{n-1} K^{(j)}_{m \,, n-1}(u) +
e^{-u+\eta(j+{1\over 2} -n)} \omega_{n} K^{(j)}_{m \,, n+1}(u)
+ {e^{\xi}\over 2 \kk \sinh \eta} e^{2 \eta (j+1-n)}K^{(j)}_{m n}(u) \,,
\non \\
\label{rel2}
\ee
where $K^{(j)}_{m n}(u)$ denotes the $(m \,, n)$ matrix element of 
$K^{(j)}(u)$. It is understood that these matrix elements vanish
for index values outside the range $[1 \,, 2j+1]$.

The relations (\ref{rel1}) and (\ref{rel2}) determine 
the matrix $K^{(j)}(u)$, up to an overall unitarization factor which 
does not concern us here. Indeed, we find that this matrix is symmetric 
$K^{(j)}_{m n}(u)=K^{(j)}_{n m}(u)$, and \footnote{Due to the presence
of the square root (which originates from the factors $\omega_{n}$
(\ref{matrixelements})), we expect that for $m \ne n$ this result is
strictly valid only for $\eta$ real.  For $\eta$ imaginary, some phase
factors may appear.}
\be
K^{(j)}_{m n}(u) &=& \kk^{n-m} 
\sqrt{\prod_{l=0}^{n-m-1}{\sinh((2j-m+1-l)\eta) \over \sinh((n-1-l)\eta)}}
\prod_{l=0}^{m-2}{\sinh((n-1-l)\eta)\over \sinh((m-1-l)\eta)}
\prod_{l=0}^{n-m-1} \sinh(2u - l \eta) \non \\
&\times& \prod_{l=0}^{2j-n} \sinh(\xi+u + (l - j + {1\over 2})\eta) 
\prod_{l=0}^{m-2} \sinh(\xi-u - (l - j + {1\over 2})\eta)\ J^{(j)}_{m n}(u) \,,
\non \\
& & \qquad m \,, n = 1 \,, 2 \,, \ldots \,, 2j+1 \,, \qquad n \ge 
m \,, 
\label{prefactor}
\ee
where the quantities $J^{(j)}_{mn}(u)$ are given by
\be
J^{(j)}_{mn}(u) = \sum_{k=0}^{\left[{2j+m-n\over 2}\right]} \kk^{2k} J^{(j\,, k)}_{mn}(u) 
\,, \qquad J^{(j\,, 0)}_{mn}(u) = 1 \,.
\label{Jj}
\ee
Finally, let us describe the quantities $J^{(j\,, k)}_{mn}(u)$ for $k \ge 1$:
for $m=1$, they are given by
\be
J^{(j\,, k)}_{1 \,, n}(u) = \sum_{l_{1}=0}^{2j-1-n}\sum_{l_{2}=l_{1}+2}^{2j-1-n}\ldots 
\sum_{l_{k}=l_{k-1}+2}^{2j-1-n} F_{l_{1}}(u \,, j \,; n)  F_{l_{2}}(u \,, j \,; n) \ldots
 F_{l_{k}}(u \,, j \,; n) \,,
\label{Jjk1}
\ee
where
\be
F_{l}(u \,, j \,; n) ={\sinh(2u-(n+l)\eta) \sinh((2j-n-l)\eta)\over 
\sinh(\xi+u+(j+{1\over 2}-n-l)\eta) \sinh(\xi+u+(j-{1\over 2}-n-l)\eta)} 
\,.
\label{Fl}
\ee
For $m \ge 2$, these quantities are determined (in terms of the
quantities with $m=1$ (\ref{Jjk1})) by the recursion relations
\be
J^{(j\,, k)}_{mn}(u) &=& a^{(j)}_{mn}(u) J^{(j\,, k)}_{m-1\,, n-1}(u) +
b^{(j)}_{mn}(u) J^{(j\,, k)}_{m-1\,, n}(u) + 
c^{(j)}_{mn}(u) J^{(j\,, k-1)}_{m-2\,, n}(u) \,, \non \\
& & \qquad m = 2\,, 3 \,, \ldots \,, 2j+1 \,, 
\label{recursion}
\ee
where
\be
a^{(j)}_{mn}(u) &=& {\sinh(\xi+u+(j-n+{3\over 2})\eta) 
\sinh(2u+ \eta)\over 
\sinh((n-m+1)\eta) \sinh(\xi-u+(j-m+{3\over 2})\eta)} \,, 
\non \\
b^{(j)}_{mn}(u) &=& -{\sinh(\xi+u+(j-m+{5\over 2})\eta) 
\sinh(2u- (n-m)\eta)\over 
\sinh((n-m+1)\eta) \sinh(\xi-u+(j-m+{3\over 2})\eta)} \,, 
\non \\
c^{(j)}_{mn}(u) &=& -{\sinh((m-2)\eta) \sinh((2j-m+3)\eta) \over 
\sinh((n-m+2)\eta) \sinh^{2}((n-m+1)\eta)} \label{coeffs}\\
&\times& {\sinh(2u+ (n-m+2)\eta) \sinh(2u- (n-m+1)\eta) \sinh(2u- (n-m)\eta)
\over 
\sinh(\xi-u+(j-m+{5\over 2})\eta)\sinh(\xi-u+(j-m+{3\over 2})\eta)}
\,. \non 
\ee 
The recursion relation (\ref{recursion}) is satisfied for $k=0$ by 
virtue of the identity
\be
1 = a^{(j)}_{mn}(u) + b^{(j)}_{mn}(u) \,.
\ee 
The recursion relations (\ref{rel1}),(\ref{rel2}) and the
expressions (\ref{prefactor})-(\ref{coeffs}) for the matrix
elements of $K^{(j)}(u)$ constitute the main results of this paper.  We
have explicitly verified for values of spin up to $j=2$ that these
results agree with those obtained by fusion \cite{MN1, Zh, Ne}, up to 
a shift of the spectral parameter and an overall factor.

It is easy to see from Eq. (\ref{prefactor}) that 
$K^{(j)}_{m n}(0)$ vanishes for $m \ne n$; and, in fact, is 
proportional to $\delta_{m n}$, as follows from also 
Eqs. (\ref{recursion}), (\ref{coeffs}). Furthermore, the
dependence of $K^{(j)}_{m n}(u)$ on the boundary parameter $\kk$
is given by $\sim \kk^{|n-m|}$, plus terms that are higher-order in $\kk$.
Hence, for $\kk=0$, $K^{(j)}_{m n}(u)$ is diagonal, and is entirely
given by Eq. (\ref{prefactor}) -- no recursion relation is then needed,
since the quantities $J^{(j\,, k)}_{m n}(u)$ do not depend on $\kk$. 
We also remark that the symmetry $K^{(j)}_{nm}(u)=K^{(j)}_{mn}(u)$ 
follows from the symmetry of the equations (\ref{rel1}),(\ref{rel2})
under transposition of $K$ and simultaneous relabeling
$n\leftrightarrow m$.  As already observed in \cite{GZ}, one can break
this symmetry and introduce a third parameter $\alpha$ into the
K-matrix by performing a change of basis $\tilde{A}(u)\mapsto
e^{i\alpha H}\tilde{A}(u)$.  While this leaves the R-matrix unchanged,
it transforms the entries of the K-matrix as $K_{mn}\mapsto
e^{i\alpha(m-n)}K_{mn}$.

It is tempting to conjecture that there exist generalizations of the 
formulas (\ref{Jjk1}), (\ref{Fl}) which are valid not just for $m=1$, but 
for all values of $m$. Indeed, we have found that an expression of the 
form
\be
J^{(j\,, k)}_{m n}(u) = \sum_{l_{1}=1-m}^{2j-1-n}\sum_{l_{2}=l_{1}+2}^{2j-1-n}
\ldots \sum_{l_{k}=l_{k-1}+2}^{2j-1-n} F_{l_{1}}(u \,, j \,; m\,, n)  
F_{l_{2}}(u \,, j \,; m\,, n) \ldots F_{l_{k}}(u \,, j \,; m\,, n) 
\label{conjec}
\ee
holds for values of $m$ up to at least $m=4$. However, we have not yet 
succeeded to find general formulas for the corresponding functions
$F_{l}(u \,, j \,; m\,, n)$.

\section{An application}\label{sec:app}

One immediate application of our result is to determine certain
coefficients appearing in the Bethe Ansatz solution of the open XXZ
quantum spin chain with nondiagonal boundary terms, defined by the
Hamiltonian \cite{Sk, dVGR}
\be
{\cal H }&=& {1\over 2}\Big\{ \sum_{n=1}^{N-1}\left( 
\sigma_{n}^{x}\sigma_{n+1}^{x}+\sigma_{n}^{y}\sigma_{n+1}^{y}
+\cosh \eta\ \sigma_{n}^{z}\sigma_{n+1}^{z}\right)\non \\
&+&\sinh \eta \Big( \coth \xi_{-} \sigma_{1}^{z}
+ {2 \kk_{-}\over \sinh \xi_{-}}\sigma_{1}^{x} 
- \coth \xi_{+} \sigma_{N}^{z}
- {2 \kk_{+}\over \sinh \xi_{+}}\sigma_{N}^{x} \Big) \Big\} \,.
\label{Hamiltonian}
\ee 
We recall \cite{Ne} that for bulk anisotropy value
\be
\eta = {i \pi\over p+1}\,, \qquad p= 1 \,, 2 \,, \ldots \,,
\label{etavalues}
\ee
(and hence $q = e^{\eta}$ is a root of unity, satisfying
$q^{p+1}=-1$), the spin-${p+1\over 2}$ transfer matrix can be
expressed in terms of a lower-spin transfer matrix, resulting in the
truncation of the fusion hierarchy.  In order to obtain this crucial
``truncation identity'' (which in turn leads to a functional relation
for the fundamental transfer matrix, and then finally to a set of
Bethe-Ansatz-like equations for the transfer-matrix eigenvalues), one
needs some knowledge of the matrix $K^{(j)}(u)$ with $j={p+1\over 2}$. 
In particular, for the $(1 \,, 1)$ matrix element, it was conjectured 
in \cite{Ne} that
\be
K^{({p+1\over 2})}_{11\  previous}(u) &\propto& n(u \,; \xi \,, \kk) 
= \sinh \left( (p+1)(\xi +u) \right)  \non \\
&+& \sum_{k=1}^{\left[{p+1\over 2}\right]}c_{p\,, k}\ 
\kk^{2k} \sinh \left( (p+1)u + (p+1 - 2k) \xi \right) \,,
\label{nfunction}
\ee 
where $c_{p\,,k}$ are some unknown coefficients. These 
coefficients were explicitly computed in \cite{Ne} for values
of $p$ up to $p=5$, and they were found to be consistent with the
formulas
\be
c_{p \,, 1} &=& p + 1 \,, \non \\
c_{p \,, 2} &=& {1\over 2}p(p-1) -1 \,.
\label{ccoefficients}
\ee
We have designated by ``previous'' the $K$ matrix appearing in \cite{Ne},
in order not to confuse it with the $K$ matrix used here, from which 
it differs by a shift of spectral parameter and an overall factor,
\be
K^{(j)}_{previous}(u) \propto K^{(j)}(u+(j-{1\over 2})\eta) \,.
\ee
Using our results (\ref{prefactor})-(\ref{Fl}) for $m=n=1$, we obtain
\be
K^{({p+1\over 2})}_{11\  previous}(u) &\propto& \sinh((p+1)(\xi + u))
\Big\{ 1 \non \\
&+& \sum_{k=1}^{\left[{p+1\over 2}\right]} \kk^{2k} 
\sum_{l_{1}=0}^{p-1}\sum_{l_{2}=l_{1}+2}^{p-1}\ldots 
\sum_{l_{k}=l_{k-1}+2}^{p-1} f_{l_{1}}(u \,;p)  f_{l_{2}}(u \,;p) 
\ldots f_{l_{k}}(u \,;p) \Big\} \,,
\label{newresult}
\ee
where
\be
f_{l}(u \,;p) &=& F_{l}(u + (j-{1\over 2})\eta \,, j \,; 1)
\Big\vert_{\eta = {i \pi\over p+1}\,, j={p+1\over 2}} \non \\ 
&=& -{\sinh(2u-(l+2)\eta) \sinh((l+1)\eta)\over 
\sinh(\xi+u-(l+2)\eta)\sinh(\xi+u-(l+1)\eta)}
\Big\vert_{\eta = {i \pi\over p+1}} \,.
\ee
Using the identity
\be
\lefteqn{\sum_{l_{1}=0}^{p-1}\sum_{l_{2}=l_{1}+2}^{p-1}\ldots 
\sum_{l_{k}=l_{k-1}+2}^{p-1} f_{l_{1}}(u \,;p)  f_{l_{2}}(u \,;p) 
\ldots f_{l_{k}}(u \,;p)}\non \\
&=&  \left( {(p+1)\over k!} \prod_{l=0}^{k-2} (p-k-l) \right)
{\sinh((p+1)u+(p+1-2k)\xi)\over \sinh((p+1)(\xi + u))} \,,
\label{identity}
\ee
and comparing (\ref{newresult}) and (\ref{identity}) with
(\ref{nfunction}), we conclude that the coefficients $c_{p \,, k}$ are
given by
\be
c_{p \,, k} = {(p+1)\over k!} \prod_{l=0}^{k-2} (p-k-l) 
 = \frac{p+1}{k}\left(\begin{array}{c}p-k\\k-1\end{array}\right)\,.
\ee
This result is evidently consistent with (\ref{ccoefficients}).

With these coefficients in hand, the Bethe-Ansatz equations can be
written down from \cite{Ne} for all the $\eta$ values (\ref{etavalues}).  
In particular, its becomes possible to study the $p \rightarrow
\infty$ limit, for which $c_{p \,, k} \sim {p^k\over k!}$.

\section{Discussion}\label{sec:disc}

We have found expressions (\ref{prefactor})-(\ref{coeffs}) for the
matrix elements of $K^{(j)}(u)$, for arbitrary spin $j$.  Since the $K$
matrix depends on two boundary parameters $\xi$ and $\kk$ as well as
the bulk anisotropy parameter $\eta$, one should {\it a priori} expect
the expression for the $K$ matrix to be more complicated than that of
the $R$ matrix (\ref{Rmatrixj}).  Our result certainly bears this out. 
Nevertheless, we expect that it may be possible to simplify our
result, perhaps along the lines of (\ref{conjec}). Indeed, a better 
understanding of the boundary quantum group symmetry may lead to a 
better choice of variables with which to express the $K$ matrix.

\section*{Acknowledgments}

One of us (RN) is grateful to P. Kulish for helpful correspondence,
and to members of the Mathematics Department of the University of York
for their warm hospitality. GWD thanks the EPSRC for an Advanced
Fellowship. The work of RN was supported in part by
the National Science Foundation under Grants PHY-9870101 and
PHY-0098088.


\begin{thebibliography}{99}
    
\bibitem{ZZ}
A.B. Zamolodchikov and Al.B. Zamolodchikov, Ann. Phys. {\it 120} (1979) 253.

\bibitem{Ba}
R.J. Baxter, {\it Exactly Solved Models in Statistical Mechanics}
(Academic Press, 1982).

\bibitem{KS0}
P.P. Kulish and E.K. Sklyanin, J. Sov. Math. {\it 19} (1982) 1596.

\bibitem{KBI}
V.E. Korepin, N.M. Bogoliubov, and A.G. Izergin, {\it Quantum Inverse 
Scattering Method, Correlation Functions and Algebraic Bethe Ansatz}
(Cambridge University Press, 1993).

\bibitem{FZ}
A.B. Zamolodchikov and V.A. Fateev, Sov. J. Nucl. Phys. {\it 32} (1980) 298.

\bibitem{KRS}
P.P. Kulish, N.Yu. Reshetikhin and E.K. Sklyanin, 
Lett. Math. Phys. {\it 5} (1981) 393.

\bibitem{KS1}
P.P. Kulish and E.K. Sklyanin, {\it Lecture Notes in Physics}, Vol. 151,
(Springer, 1982) 61.

\bibitem{KR}
P.P. Kulish and N.Yu. Reshetikhin, J. Sov. Math. {\it 23} (1983) 2435.

\bibitem{Dr}
V.G. Drinfel'd, J. Sov. Math. {\it 41} (1988) 898.

\bibitem{Ji}
M. Jimbo, Int. J. Mod. Phys. {\it A4} (1989) 3759.

\bibitem{CP}
V. Chari and A. Pressley, {\it A guide to quantum groups}, 
(Cambridge University Press, 1994).

\bibitem{Za1}
A.B. Zamolodchikov, ``Fractional-spin integrals of motion in perturbed
conformal field theory,'' in {\it Fields, Strings and Quantum Gravity}, eds.
H. Guo, Z. Qiu and H. Tye, (Gordon and Breach, 1989).

\bibitem{BL} 
D. Bernard and A. LeClair, Commun. Math. Phys. {\it 142} (1991) 99.

\bibitem{Ch} 
I.V. Cherednik, Theor. Math. Phys. {\it 61} (1984) 977.

\bibitem{Sk}
E.K. Sklyanin, J. Phys. {\it A21} (1988) 2375.

\bibitem{GZ}
S. Ghoshal and A.B. Zamolodchikov, Int. J. Mod. Phys. {\it A9} (1994)
3841.

\bibitem{dVGR}
H.J. de Vega and A. Gonz\'alez-Ruiz, J. Phys. {\it A26} (1993) L519.

\bibitem{IOZ}
T. Inami, S. Odake and Y.-Z. Zhang, Nucl. Phys. {\it B470} (1996) 419.

\bibitem{MNR}
L. Mezincescu, R.I. Nepomechie and V. Rittenberg, 
Phys. Lett. {\it A147} (1990) 70.

\bibitem{MN1}
L. Mezincescu and R.I. Nepomechie, J. Phys. {\it A25} (1992) 2533.

\bibitem{Zh}
Y.-K. Zhou, Nucl. Phys. {\it B458} (1996) 504.

\bibitem{KS2}
P.P. Kulish and E.K. Sklyanin, J. Phys. {\it A25} (1992) 5963.

\bibitem{KuSa}
P.P. Kulish and R. Sasaki, Prog. Theor. Phys. {\it 89} (1993) 741. 

\bibitem{CG}
E. Cremmer and J.-L. Gervais, Commun. Math. Phys. {\it 144} (1992) 279. 

\bibitem{MN2}
L. Mezincescu and R.I. Nepomechie, Int. J. Mod. Phys. {\it A13} (1998) 
2747.

\bibitem{DM}
G.W. Delius and N.J. MacKay, ``Quantum group symmetry in sine-Gordon 
and affine Toda field theories on the half-line,'' {\tt hep-th/0112023}.

\bibitem{Ne}
R.I. Nepomechie, Nucl. Phys. {\it B622} (2002) 615.


    
\end{thebibliography}
\end{document}